\documentclass[pdflatex,sn-mathphys-num]{sn-jnl}

\usepackage{graphicx}%
\usepackage{multirow}%
\usepackage{amsmath,amssymb,amsfonts}%
\usepackage{amsthm}%
\usepackage{mathrsfs}%
\usepackage[title]{appendix}%
\usepackage{xcolor}%
\usepackage{textcomp}%
\usepackage{manyfoot}%
\usepackage{booktabs}%
\usepackage{algorithm}%
\usepackage{algorithmicx}%
\usepackage{algpseudocode}%
\usepackage{listings}%
\usepackage{lscape} 

\theoremstyle{thmstyleone}%
\theoremstyle{thmstyletwo}%

\theoremstyle{thmstylethree}%

\begin{document}

\title[Article Title]{Configurable controlled teleportation using multipartite GHZ states}


\author[1]{\fnm{Soubhik} \sur{De}}

\author[1]{\fnm{Vedhanayagi} \sur{R.}}

\author[1]{\fnm{S.V.M.} \sur{Satyanarayana}}

\author*[1,]{\fnm{Alok} \sur{Sharan}}\email{aloksharan@pondiuni.ac.in}

\affil[1]{\orgdiv{Department of Physics}, \orgname{Pondicherry University}, \orgaddress{\street{Kalapet}, \city{Pondicherry}, \postcode{605014}, \state{Puducherry}, \country{India}}}


\abstract{We propose a controlled quantum teleportation protocol for securely transferring an unknown $n$-qubit state from a sender to a receiver, under the supervision of $m$ controller participants. The protocol uses $n$ copies of an $m$-qubit Greenberger-Horne-Zeilinger state as the quantum resource.
Message qubits can be distributed among participants to enhance security against targeted external attacks. Each intermediate party may hold at most one resource qubit, reducing the total number of resource qubits required. The sender selects the end receiver during protocol execution, ensuring anonymity and minimizing the risk of interception by an external eavesdropper.
We assess the protocol's performance by calculating teleportation fidelities for various $m$ and $n$ values and visualize the quantum states through Hinton diagrams. The results confirm the protocol's effectiveness for secure quantum communication in multi-party settings.}

\keywords{Multi-qubit Teleportation, Configurable Protocol, Anonymous End Receiver, Distributed Message Qubits}

\maketitle

\section{Introduction}\label{sec_1}

Quantum Teleportation utilizes the features of entanglement and superposition to transfer quantum information stored in quantum states from the sender (usually referred to as Alice) to the receiver (Bob) \cite{rivest1978}, with the assistance of local operations and classical communications. This transfer of information strikingly does not involve any physical movement of the participating systems. The first quantum teleportation protocol was proposed in 1993 by Bennett et al. \cite{bennett1993}, which was experimentally implemented by Bouwmeester et al. \cite{bouwmeester1997} four years later.

The field of quantum teleportation has since seen numerous advancements and variations tailored to accommodate multi-qubit systems \cite{cao2007,li2016} and networked quantum nodes \cite{pirandola2015}. With the development of large-scale quantum networks \cite{allati2011,wei2022} and the envisioned ``quantum internet" \cite{kimble2008}, the ability to securely and efficiently transmit quantum information across multiple nodes has become a critical objective \cite{sisodia2017,hofmann2019}.

Distributed quantum computing \cite{caleffi2024}, demands not only scalable qubit resources but also resilient teleportation schemes that maintain high fidelity even in the presence of noise in the channel \cite{jung2008,adhikari2012}. Such protocols use multipartite entangled states as resource to support teleportation across several participants \cite{briegel1998,zwerger2012}.

Advances in teleportation protocols have led to the development of controlled and bidirectional teleportation schemes. In controlled teleportation, one or more participants act as controllers whose cooperation is essential for faithful teleportation. Karlsson et al. \cite{karlsson1998} introduced an early model of controlled quantum teleportation utilizing GHZ states, and further studies have improved the security and reliability of multi-party quantum communication \cite{yan2005,song2008,wang2009,guo2011,li2013,kumar2020,yang2023,kirdi2023,yuan2024,yuan2024b}. Bidirectional teleportation allows for the mutual transmission of quantum information between two parties, effectively doubling communication capacity, while further works have improved security \cite{duan2014,hassanpour2016,choudhury2018,zhou2019,zhou2019b,verma2021,wang2022,verma2020}. Building on these methods, bidirectional controlled teleportation protocols have been developed, combining the features of controlled and bidirectional teleportation to enable secure, reciprocal quantum communication under the oversight of controller participants \cite{chou2014,chen2015,jiang2020,du2020,verma2021b,yang2023b}. The development and standardization of a multi-nodal protocol \cite{wen2010,choudhury2018b} for distributed quantum computing is necessary for the efficient utilization of quantum networks for secure quantum communication.

In this work, we propose a multi-nodal protocol that teleports an $n$-qubit state via $m$ controllers, where we distribute both the message qubits and resource qubits in various configurations according to use cases. In Section \ref{sec_2}, we first demonstrate the protocol for 3 participants, then showcase the generalization of the end receiver unitary operations which lets us accommodate $m$ controllers. In Section \ref{sec_3}, we extend the protocol to teleport a general $n$-qubit state through $m$ participants. Section \ref{sec_5} details the obtained fidelities of teleportation in ideal simulations of the teleportation protocol and showcase Hinton diagrams of input and teleported states for different $m$ and $n$ values of the protocol. Section \ref{sec_6} ends with the conclusion.

\section{Control of teleportation by \textit{m} participants}
\label{sec_2}
We begin by examining the teleportation protocol with three participants, extend the analysis to four participants, and subsequently generalize the framework to accommodate $m$ participants.

\subsection{1-qubit Controlled Teleportation via 3 participants}
\label{sec_2.1}
The three participants Alice, Charlie and Bob initially share the teleportation resource, a maximally-entangled 3-qubit GHZ state given in Equation \ref{eqn_1}.
\begin{equation}
    |\text{GHZ}\rangle_{q_1 q_2 q_3} = \frac{1}{\sqrt{2}} (|000\rangle + |111\rangle)_{q_1 q_2 q_3}
    \label{eqn_1}
\end{equation}
where qubit $q_1$ is with sender Alice, qubit $q_2$ is with the intermediate participant Charlie and qubit $q_3$ is with receiver Bob. Alice also has the unknown 1-qubit state $|\psi^1\rangle$
given in Equation \ref{eqn_2}.
\begin{equation}
|\psi^1\rangle_{q_0} = (\alpha|0\rangle + \beta|1\rangle)_{q_0}
\label{eqn_2}
\end{equation}
where $|\alpha|^2 + |\beta|^2 = 1$. The composite system state is given below in Equation \ref{eqn_3}.
\begin{align}
|\psi^1\rangle_{q_0} \otimes |\text{GHZ}\rangle_{q_1 q_2 q_3} &= (\alpha |0\rangle + \beta |1\rangle)_{q_0} \otimes \frac{1}{\sqrt{2}} (|000\rangle + |111\rangle)_{q_1 q_2 q_3} \nonumber \\
&= \frac{1}{\sqrt{2}} (\alpha |0000\rangle + \alpha |0111\rangle + \beta |1000\rangle + \beta |1111\rangle)_{q_0 q_1 q_2 q_3}
\label{eqn_3}
\end{align}

Alice now does a Bell measurement (BM) on her qubits $q_0$ and $q_1$, collapsing the system into the post-measurement state $|\psi^{1'}\rangle$. Alice saves her results in classical bits $c_0$ and $c_1$ respectively. Charlie and Bob's post-measurement states are listed in Table \ref{tab_1}.

\begin{table}[b]
\centering
\caption{\label{tab_1}Post-measurement states left with Charlie and Bob after Alice's BMs.}
\begin{tabular}{@{}ll@{}}
\toprule
Alice's measurement results $(c_0,c_1)$ & Charlie and Bob's state $|q_2q_3\rangle$\\
\midrule
00 & $\alpha|00\rangle + \beta|11\rangle$ \\
01 & $\alpha|11\rangle + \beta|00\rangle$ \\
10 & $\alpha|00\rangle - \beta|11\rangle$ \\
11 & $\alpha|11\rangle - \beta|00\rangle$ \\
\bottomrule
\end{tabular}
\end{table}

\begin{figure}
    \centering
    \includegraphics[width=0.8\columnwidth]{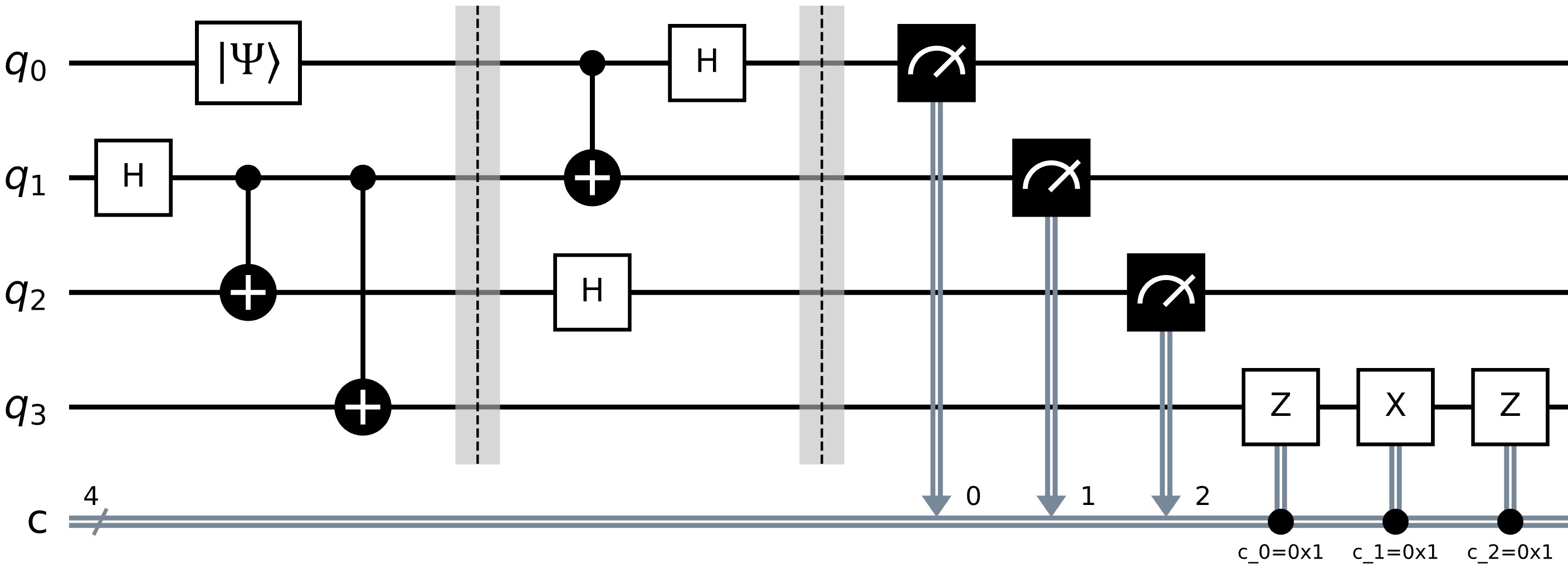}
    \caption{\label{fig_1}Quantum circuit illustrating the protocol for $m=3$ and $n=1$ configuration. Qubit $q_0$ initially contains the unknown message state $|\psi^1\rangle$, while qubits $q_1$ to $q_3$ initially contain the 3-qubit GHZ state, and $c$ is a 3-bit classical register.}
\end{figure}

The intermediate participant Charlie now applies a Hadamard gate on his qubit $q_2$, measures $q_2$ in Z basis and saves this result in classical bit $c_2$. The parties then send all classical bits to Bob, using which he applies unitary rotations (UR) on his qubit $q_3$, hence reconstructing the unknown 1-qubit state  $|\psi^1\rangle$. All possible UR are listed in Table \ref{tab_2}.
\begin{table}
\centering
\caption{\label{tab_2}Bob's unitary rotations using participants' measurement values.}
\begin{tabular}{@{}llll@{}}
\toprule
Alice's measurement & Charlie's measurement & Bob's qubit & Bob's \\
results $(c_0,c_1)$ & results $(c_2)$ & state $|q_3\rangle$ & rotations\\
\midrule
\multirow{2}{*}{00} & 0 & $\alpha|0\rangle + \beta|1\rangle$ & I \\
& 1 & $\alpha|0\rangle - \beta|1\rangle$ & Z \\
\midrule
\multirow{2}{*}{01} & 0 & $\alpha|1\rangle + \beta|0\rangle$ & X \\
& 1 & $-\alpha|1\rangle + \beta|0\rangle$ & XZ \\
\midrule
\multirow{2}{*}{10} & 0 & $\alpha|0\rangle - \beta|1\rangle$ & Z \\
& 1 & $\alpha|0\rangle + \beta|1\rangle$ & I \\
\midrule
\multirow{2}{*}{11} & 0 & $\alpha|1\rangle - \beta|0\rangle$ & ZX \\
& 1 & $-\alpha|1\rangle - \beta|0\rangle$ & ZXZ \\
\bottomrule
\end{tabular}
\end{table}
Bob's UR can be summarized using the formula below in Equation \ref{eqn_9}. The circuit for $m=3$ and $n=1$ is shown in Figure \ref{fig_1}.
\begin{equation}
    \text{Unitary rotations applied by Bob on qubit } q_3 \implies Z^{c_0} \cdot X^{c_1} \cdot Z^{c_2}
    \label{eqn_9}
\end{equation}
\subsection{1-qubit Controlled Teleportation via 4 participants}\label{sec_2.2}
\begin{figure}
    \centering
    \includegraphics[width=\columnwidth]{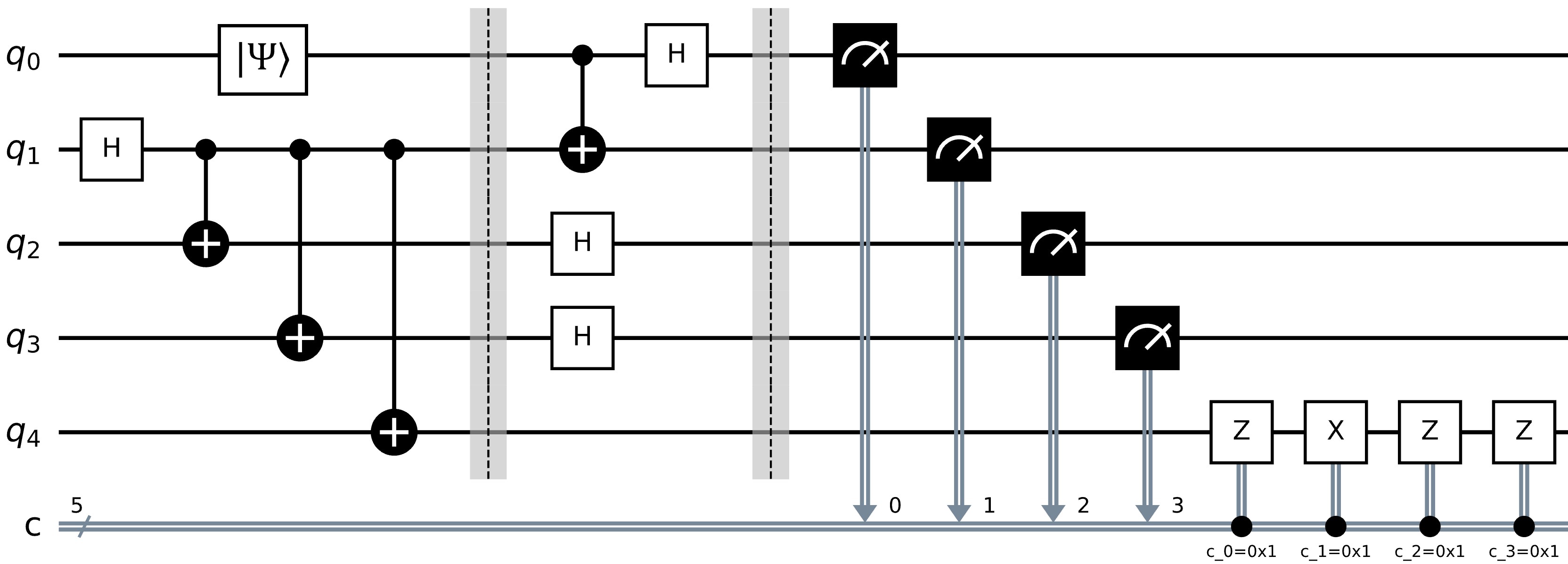}
    \caption{\label{fig_2}Quantum circuit illustrating the protocol for $m=4$ and $n=1$ configuration. Qubit $q_0$ initially contains the unknown message state $|\psi^1\rangle$, while qubits $q_1$ to $q_4$ initially contain the 4-qubit $|4\text{GHZ}\rangle$ state, and $c$ is a 4-bit classical register.}
\end{figure}

The four participants Alice, $\text{Charlie}_0$, $\text{Charlie}_1$ and Bob share the 4-qubit GHZ state given in Equation \ref{eqn_10}.
\begin{equation}
    |4\text{GHZ}\rangle = \frac{1}{\sqrt{2}}(|0000\rangle + |1111\rangle)_{q_1q_2q_3q_4}
    \label{eqn_10}
\end{equation}
where qubit $q_1$ is with Alice, $q_2$ is with $\text{Charlie}_0$, $q_3$ is with $\text{Charlie}_1$ and $q_4$ is with Bob. Alice also has the qubit $q_0$ which contains the unknown 1-qubit state $|\psi^1\rangle$, given in Equation \ref{eqn_2}. The composite system state is hence given in Equation \ref{eqn_11}.
\begin{align}
    \label{eqn_11}
    |\psi^1\rangle \otimes|4\text{GHZ}\rangle &= (\alpha|0\rangle + \beta|1\rangle)_{q_0}\otimes\frac{1}{\sqrt{2}}(|0000\rangle+|1111\rangle)_{q_1q_2q_3q_4} \nonumber \\
    &= \frac{1}{\sqrt{2}}[\alpha(|00000\rangle + |01111\rangle) + \beta(|10000\rangle + |11111\rangle)]_{q_0q_1q_2q_3q_4}
\end{align}
Alice performs BMs on her qubits $q_0$ and $q_1$ and saves the measurement results in $c_0$ and $c_1$. The possible post-measurement states are listed in Table \ref{tab_3}.
\begin{table}
    \centering
    \caption{\label{tab_3}Post-measurement states left with $\text{Charlie}_0$, $\text{Charlie}_1$ and Bob after Alice's BMs.}
    \begin{tabular}{@{}ll@{}}
    \toprule
    Alice's measurement results $(c_0,c_1)$ & Post-measurement state $|q_2q_3q_4\rangle$ \\
    \midrule
    00 & $\alpha|000\rangle+\beta|111\rangle$ \\
    01 & $\alpha|111\rangle+\beta|000\rangle$ \\
    10 & $\alpha|000\rangle-\beta|111\rangle$ \\
    11 & $\alpha|111\rangle-\beta|000\rangle$ \\
    \bottomrule
    \end{tabular}
\end{table}
Intermediate participants $\text{Charlie}_0$ and $\text{Charlie}_1$ each apply a Hadamard gate on $q_2$ and $q_3$, measure in Z basis and save results into $c_2$ and $c_3$ respectively. The parties then send all classical bits to Bob, using which he applies UR on his qubit $q_4$, reconstructing the unknown 1-qubit state  $|\psi^1\rangle$. All possible rotations are listed in Table \ref{tab_4}. Bob's UR can be summarized using the formula below in Equation \ref{eqn_15}. The  circuit for $m=4$ and $n=1$ is shown in Figure \ref{fig_2}.
\begin{equation}
    \label{eqn_15}
    \text{Unitary rotations applied by Bob on qubit } q_4 \implies Z^{c_0} \cdot X^{c_1} \cdot Z^{c_2} \cdot Z^{c_3}
\end{equation}
\begin{table}[h]
    \centering
    \caption{\label{tab_4}Bob's unitary rotations using participants' measurement values.}
    \begin{tabular}{@{}llll@{}}
    \toprule
    Alice's measurement & $\text{Chalrie}_0$ and $\text{Chalrie}_1$'s & Bob's qubit & Bob's \\
    results $(c_0,c_1)$ & measurement results $(c_2,c_3)$ & state $|q_4\rangle$ & rotations \\
    \midrule
    \multirow{4}{*}{00} & 00 & $\alpha|0\rangle+\beta|1\rangle$ & I \\
     & 01 & $\alpha|0\rangle-\beta|1\rangle$ & Z \\
     & 10 & $\alpha|0\rangle-\beta|1\rangle$ & Z \\
     & 11 & $\alpha|0\rangle+\beta|1\rangle$ & I \\
     \midrule
     \multirow{4}{*}{01} & 00 & $\alpha|1\rangle+\beta|0\rangle$ & X \\
     & 01 & $-\alpha|1\rangle+\beta|0\rangle$ & XZ \\
     & 10 & $-\alpha|1\rangle+\beta|0\rangle$ & XZ \\
     & 11 & $\alpha|1\rangle+\beta|0\rangle$ & X \\
     \midrule
     \multirow{4}{*}{10} & 00 & $\alpha|0\rangle-\beta|1\rangle$ & Z \\
     & 01 & $\alpha|0\rangle+\beta|1\rangle$ & I \\
     & 10 & $\alpha|0\rangle+\beta|1\rangle$ & I \\
     & 11 & $\alpha|0\rangle-\beta|1\rangle$ & Z \\
     \midrule
     \multirow{4}{*}{11} & 00 & $\alpha|1\rangle-\beta|0\rangle$ & ZX \\
     & 01 & $-\alpha|1\rangle-\beta|0\rangle$ & ZXZ \\
     & 10 & $-\alpha|1\rangle-\beta|0\rangle$ & ZXZ \\
     & 11 & $\alpha|1\rangle-\beta|0\rangle$ & ZX \\
     \bottomrule
    \end{tabular}
\end{table}

\subsection{1-qubit Controlled Teleportation via $m$ Participants}

In this generalized protocol, there are $(m-2)$ intermediate participants $\text{Charlie}_i$ where $i=0,1,2,..,(m-3)$, excluding the sender Alice and the receiver Bob. The generalized maximally-entangled $m$-qubit GHZ state shared by these parties is given below in Equation \ref{eqn_16}.
\begin{align}
    \label{eqn_16}
    |m\text{GHZ}\rangle_{q_1q_2q_3..q_m} &= \frac{1}{\sqrt{2}}(|0\rangle^{\otimes m})+|1\rangle^{\otimes m})_{q_1q_2q_3..q_m} \nonumber \\
    &= \frac{1}{\sqrt{2}}(|0..m\text{ times}\rangle+|1..m\text{ times}\rangle)_{q_1q_2q_3..q_m}
\end{align}
where $q_1$ is with Alice, $q_m$ is with Bob and the intermediate qubits are with the $\text{Charlie}_i$'s. Additionally, the sender Alice has an unknown 1-qubit state $|\psi^1\rangle$ given in Equation \ref{eqn_2}. The composite system state is given below in Equation \ref{eqn_17}.
\begin{align}
    \label{eqn_17}
    |\psi^1\rangle_{q_0} \otimes|m\text{GHZ}\rangle_{q_1q_2q_3..q_m} &= (\alpha|0\rangle+\beta|1\rangle)_{q_0}\otimes\frac{1}{\sqrt{2}}(|0\rangle^{\otimes m}+|1\rangle^{\otimes m})_{q_1q_2q_3..q_m} \nonumber \\
    &= \frac{1}{\sqrt{2}}
    \left[
    \begin{aligned}
        &\alpha|0\rangle\otimes|0\rangle^{\otimes m}+\alpha|0\rangle\otimes|1\rangle^{\otimes m} + \\
        &\beta|1\rangle\otimes|0\rangle^{\otimes m}+\beta|1\rangle\otimes|1\rangle^{\otimes m} \\
    \end{aligned}
    \right]_{q_0q_1q_2q_3..q_m}
\end{align}
Alice now performs BMs on her qubits $q_0$ and $q_1$ and saves the results in $c_0$ and $c_1$ respectively. All possible measurement results and corresponding post-measurement states are listed in Table \ref{tab_5}.
\begin{table}
    \centering
    \caption{\label{tab_5}Post-measurement states left with $\text{Charlie}_i$'s and Bob after Alice's BM.}
    \begin{tabular}{@{}ll@{}}
    \toprule
    Alice's measurement results $(c_0,c_1)$ & Post-measurement state $|q_2q_3..q_m\rangle$ \\
    \midrule
    00 & $\alpha|0\rangle^{\otimes(m-1)}+\beta|1\rangle^{\otimes(m-1)}$ \\
    01 & $\alpha|1\rangle^{\otimes(m-1)}+\beta|1\rangle^{\otimes(m-1)}$ \\
    10 & $\alpha|0\rangle^{\otimes(m-1)}-\beta|1\rangle^{\otimes(m-1)}$ \\
    11 & $\alpha|1\rangle^{\otimes(m-1)}-\beta|1\rangle^{\otimes(m-1)}$ \\
    \bottomrule
    \end{tabular}
\end{table}

The intermediate participants $\text{Charlie}_i$'s each apply a Hadamard gate to their corresponding qubits, measure in the Z basis and save the results in corresponding classical bits $c_{(i+2)}$. The resulting unitary rotations to be implemented by Bob can be summarized below in Equation \ref{eqn_20}.
\begin{align}
    \label{eqn_20}
    \text{Unitary rotations applied by Bob on qubit } q_m &\implies Z^{c_0} \cdot X^{c_1} \cdot Z^{c_2} \cdot Z^{c_3} \cdot ..\cdot Z^{c_{(m-1)}} \nonumber \\
    & \implies Z^{c_0} \cdot X^{c_1} \cdot Z^{[c_2\oplus c_3\oplus..\oplus c_{(m-1)}]}
\end{align}
where $\oplus$ represents the binary \textbf{XOR} operation between classical bits. 

\section{Teleportation of \textit{n} qubits}
\label{sec_3}
The message state is generalized to an unknown general $n$-qubit state $|\psi^n\rangle$, while the teleportation channel is changed to a product state formed from $n$ instances of $m$-qubit GHZ states, given in Equation \ref{eqn_30}. $(m\times n)$ resource qubits are used for the protocol due to these modifications. We begin by examining the teleportation protocol with 2 qubits, and subsequently generalize the framework to accommodate $n$ qubits.

\subsection{2-qubit Controlled Teleportation via 3 participants}
\label{sec_3.1}
The number of qubits $(n)$ in the message state is 2, given below by the unknown state $|\psi^2\rangle$ in Equation \ref{eqn_21}.
\begin{equation}
    \label{eqn_21}
    |\psi^2\rangle = (\alpha|00\rangle+\beta|01\rangle+\gamma|10\rangle+\delta|11\rangle)_{q_0q_1}
\end{equation}
where $|\alpha|^2 + |\beta|^2 + |\gamma|^2 +|\delta|^2 = 1$. The resource is a 6-qubit state $|2,3\text{GHZ}\rangle$ given below in Equation \ref{eqn_22}.
\begin{align}
    \label{eqn_22}
    |2,3\text{GHZ}\rangle &= \frac{1}{2}[(|000\rangle+|111\rangle)_{q_2q_4q_6}\otimes(|000\rangle+|111\rangle)_{q_3q_5q_7}] \nonumber \\
    &= \frac{1}{2}[|000000\rangle+|000111\rangle+|111000\rangle+|111111\rangle]_{q_2q_4q_6q_3q_5q_7} \nonumber \\
    &= \frac{1}{2}[|000000\rangle+|010101\rangle+|101010\rangle+|111111\rangle]_{q_2q_3q_4q_5q_6q_7}
\end{align}
It is a product state of two separate 3-qubit maximally-entangled GHZ states, where adjacent qubits are given to adjacent participants, i.e., Alice gets qubits $q_2$ and $q_3$, Bob gets qubits $q_6$ and $q_7$ and intermediate participant Charlie get qubits $q_4$ and $q_5$. The composite system state is given below in Equation \ref{eqn_23}.
\begin{align}
    \label{eqn_23}
    &|\psi^2\rangle\otimes |2,3\text{GHZ}\rangle \nonumber \\[2pt]
    &= (\alpha|00\rangle+\beta|01\rangle+\gamma|10\rangle+\delta|11\rangle)_{q_0q_1} \otimes \nonumber \\
    &\hspace{1.3in} \frac{1}{2}[|000000\rangle+|010101\rangle+|101010\rangle+|111111\rangle]_{q_2q_3q_4q_5q_6q_7} \\ \nonumber
\end{align}
\begin{figure}
    \centering
    \includegraphics[width=0.99\columnwidth]{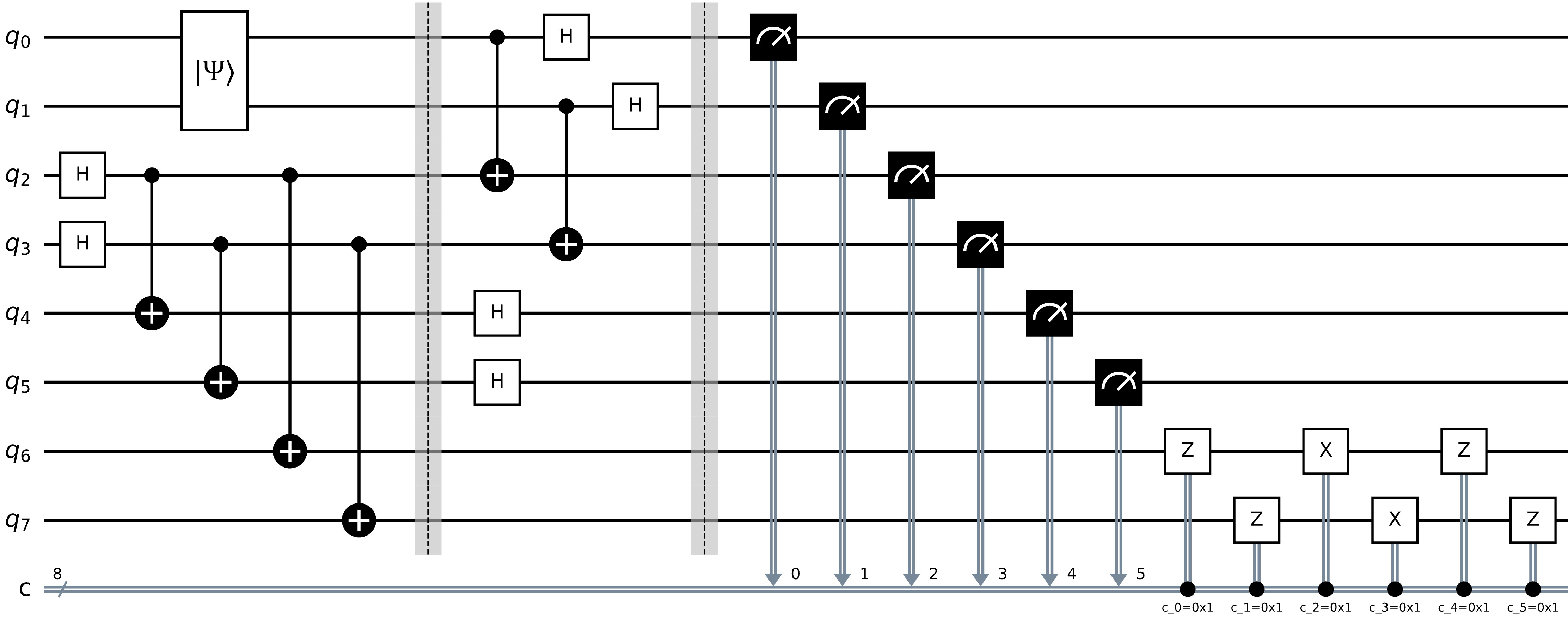}
    \caption{\label{fig_3}Quantum circuit illustrating the protocol for $m=3$ and $n=2$ configuration. Qubits $q_0$ and $q_1$ initially contains the unknown message state $|\psi^2\rangle$, while qubits $q_2$ to $q_7$ initially contain the 6-qubit resource state, and $c$ is a 6-bit classical register.}
\end{figure}
\vspace{-40pt}

Alice performs BMs on pairs of her qubits, where each pair consists of a message qubit and a corresponding qubit from the resource state, viz., $BM_{q_0,q_2}$ and $BM_{q_1,q_3}$, and saves the results in $c_0$, $c_1$, $c_2$ and $c_3$. The possible post-measurement states are listed in Table \ref{tab_6}. Intermediate participant Charlie applies a Hadamard gate on each of their qubits, measure them in Z basis and save these measurements in $c_4$ and $c_5$. Bob applies UR on each of his qubits $q_6$ and $q_7$, reconstructing the unknown 2-qubit state $|\psi^2\rangle$. All possible rotations are listed in Table \ref{tab_7}, and these can summarized below using the formula in Equation \ref{eqn_27}. The circuit for $m=3$ and $n=2$ is given in Figure \ref{fig_3}.
\begin{align}
    \label{eqn_27}
    &\text{Unitary rotations applied by Bob on qubit } q_6 \implies Z^{c_0}\cdot X^{c_2}\cdot Z^{c_4} \nonumber \\
    &\text{Unitary rotations applied by Bob on qubit } q_7 \implies Z^{c_1}\cdot X^{c_3}\cdot Z^{c_5}
\end{align}
\vspace{-20pt}

\begin{table}[h]
    \caption{\label{tab_6}Post-measurement states left with Charlie and Bob after Alice's BMs.}
    \centering
    \begin{tabular}{@{}ll@{}}
    \toprule
    Alice's measurement results $(c_0, c_1, c_2, c_3)$  & Post-measurement state $|q_4q_5q_6q_7\rangle$\\
    \midrule
    0000 & $\alpha|0000\rangle+\beta|0101\rangle+\gamma|1010\rangle+\delta|1111\rangle$ \\
    0001 & $\alpha|0101\rangle+\beta|0000\rangle+\gamma|1111\rangle+\delta|1010\rangle$ \\
    0010 & $\alpha|1010\rangle+\beta|1111\rangle+\gamma|0000\rangle+\delta|0101\rangle$ \\
    0011 & $\alpha|1111\rangle+\beta|1010\rangle+\gamma|0101\rangle+\delta|0000\rangle$ \\
    0100 & $\alpha|0000\rangle-\beta|0101\rangle+\gamma|1010\rangle-\delta|1111\rangle$ \\
    0101 & $\alpha|0101\rangle-\beta|0000\rangle+\gamma|1111\rangle-\delta|1010\rangle$ \\
    0110 & $\alpha|1010\rangle-\beta|1111\rangle+\gamma|0000\rangle-\delta|0101\rangle$ \\
    0111 & $\alpha|1111\rangle-\beta|1010\rangle+\gamma|0101\rangle-\delta|0000\rangle$ \\
    1000 & $\alpha|0000\rangle+\beta|0101\rangle-\gamma|1010\rangle-\delta|1111\rangle$ \\
    1001 & $\alpha|0101\rangle+\beta|0000\rangle-\gamma|1111\rangle-\delta|1010\rangle$ \\
    1010 & $\alpha|1010\rangle+\beta|1111\rangle-\gamma|0000\rangle-\delta|0101\rangle$ \\
    1011 & $\alpha|1111\rangle+\beta|1010\rangle-\gamma|0101\rangle-\delta|0000\rangle$ \\
    1100 & $\alpha|0000\rangle-\beta|0101\rangle-\gamma|1010\rangle+\delta|1111\rangle$ \\
    1101 & $\alpha|0101\rangle-\beta|0000\rangle-\gamma|1111\rangle+\delta|1010\rangle$ \\
    1110 & $\alpha|1010\rangle-\beta|1111\rangle-\gamma|0000\rangle+\delta|0101\rangle$ \\
    1111 & $\alpha|1111\rangle-\beta|1010\rangle-\gamma|0101\rangle+\delta|0000\rangle$ \\
    \bottomrule
    \end{tabular}
\end{table}

\begin{landscape}
    \begin{table}
        \centering
        \caption{\label{tab_7}Bob's unitary rotations using Alice's measurement results $(c_0,c_1,c_2,c_3)$ and Charlie's measurement results $(c_4,c_5)$}
        \begin{tabular}{@{}llll|llll@{}}
            \toprule
            Alice's & Chalrie's & Bob's & Bob's & Alice's & Chalrie's & Bob's & Bob's \\
            results  & results  & state $|q_6\rangle$ & rotations & results  & results  & state $|q_6\rangle$ & rotations \\
            \midrule
            \multirow{4}{*}{0000} & 00 & $\alpha|00\rangle+\beta|01\rangle+\gamma|10\rangle+\delta|11\rangle$ & $I\otimes I$
            & \multirow{4}{*}{1000} & 00 & $\alpha|00\rangle+\beta|01\rangle-\gamma|10\rangle-\delta|11\rangle$ & $Z\otimes I$\\
            & 01 & $\alpha|00\rangle-\beta|01\rangle+\gamma|10\rangle-\delta|11\rangle$ & $I\otimes Z$ & 
            & 01 & $\alpha|00\rangle-\beta|01\rangle-\gamma|10\rangle+\delta|11\rangle$ & $Z\otimes Z$ \\
            & 10 & $\alpha|00\rangle+\beta|01\rangle-\gamma|10\rangle-\delta|11\rangle$ & $Z\otimes I$ & 
            & 10 & $\alpha|00\rangle+\beta|01\rangle+\gamma|10\rangle+\delta|11\rangle$ & $I\otimes I$ \\
            & 11 & $\alpha|00\rangle-\beta|01\rangle-\gamma|10\rangle+\delta|11\rangle$ & $Z\otimes Z$ & 
            & 11 & $\alpha|00\rangle-\beta|01\rangle+\gamma|10\rangle-\delta|11\rangle$ & $I\otimes Z$ \\
            \midrule
            \multirow{4}{*}{0001} & 00 & $\alpha|01\rangle+\beta|00\rangle+\gamma|11\rangle+\delta|10\rangle$ & $I\otimes X$ 
            & \multirow{4}{*}{1001} & 00 & $\alpha|01\rangle+\beta|00\rangle-\gamma|11\rangle-\delta|10\rangle$ & $Z\otimes X$ \\
            & 01 & $\alpha|01\rangle-\beta|00\rangle+\gamma|11\rangle-\delta|10\rangle$ & $I\otimes XZ$ & 
            & 01 & $\alpha|01\rangle-\beta|00\rangle-\gamma|11\rangle+\delta|10\rangle$ & $Z\otimes XZ$ \\
            & 10 & $\alpha|01\rangle+\beta|00\rangle-\gamma|11\rangle-\delta|10\rangle$ & $Z\otimes X$ & 
            & 10 & $\alpha|01\rangle+\beta|00\rangle+\gamma|11\rangle+\delta|10\rangle$ & $I\otimes X$ \\
            & 11 & $\alpha|01\rangle-\beta|00\rangle-\gamma|11\rangle+\delta|10\rangle$ & $Z\otimes XZ$ & 
            & 11 & $\alpha|01\rangle-\beta|00\rangle+\gamma|11\rangle-\delta|10\rangle$ & $I\otimes XZ$ \\
            \midrule
            \multirow{4}{*}{0010} & 00 & $\alpha|10\rangle+\beta|11\rangle+\gamma|00\rangle+\delta|01\rangle$ & $X\otimes I$ 
            & \multirow{4}{*}{1010} & 00 & $-\alpha|10\rangle-\beta|11\rangle+\gamma|00\rangle+\delta|01\rangle$ & $ZX\otimes I$ \\
            & 01 & $\alpha|10\rangle-\beta|11\rangle+\gamma|00\rangle-\delta|01\rangle$ & $X\otimes Z$ & 
            & 01 & $-\alpha|10\rangle+\beta|11\rangle+\gamma|00\rangle-\delta|01\rangle$ & $ZX\otimes Z$ \\
            & 10 & $\alpha|10\rangle+\beta|11\rangle-\gamma|00\rangle-\delta|01\rangle$ & $XZ\otimes I$ & 
            & 10 & $-\alpha|10\rangle-\beta|11\rangle-\gamma|00\rangle-\delta|01\rangle$ & $ZXZ\otimes I$ \\
            & 11 & $\alpha|10\rangle-\beta|11\rangle-\gamma|00\rangle+\delta|01\rangle$ & $XZ\otimes Z$ & 
            & 11 & $-\alpha|10\rangle+\beta|11\rangle-\gamma|00\rangle+\delta|01\rangle$ & $ZXZ\otimes Z$ \\
            \midrule
            \multirow{4}{*}{0011} & 00 & $\alpha|11\rangle+\beta|10\rangle+\gamma|01\rangle+\delta|00\rangle$ & $X\otimes X$ 
            & \multirow{4}{*}{1011} & 00 & $-\alpha|11\rangle-\beta|10\rangle+\gamma|01\rangle+\delta|00\rangle$ & $ZX\otimes X$ \\
            & 01 & $\alpha|11\rangle-\beta|10\rangle+\gamma|01\rangle-\delta|00\rangle$ & $X\otimes XZ$ & 
            & 01 & $-\alpha|11\rangle+\beta|10\rangle+\gamma|01\rangle-\delta|00\rangle$ & $ZX\otimes XZ$ \\
            & 10 & $\alpha|11\rangle+\beta|10\rangle-\gamma|01\rangle-\delta|00\rangle$ & $XZ\otimes X$ & 
            & 10 & $-\alpha|11\rangle-\beta|10\rangle-\gamma|01\rangle-\delta|00\rangle$ & $ZXZ\otimes X$ \\
            & 11 & $\alpha|11\rangle-\beta|10\rangle-\gamma|01\rangle+\delta|00\rangle$ & $XZ\otimes XZ$ & 
            & 11 & $-\alpha|11\rangle+\beta|10\rangle-\gamma|01\rangle+\delta|00\rangle$ & $ZXZ\otimes XZ$ \\
            \midrule
            \multirow{4}{*}{0100} & 00 & $\alpha|00\rangle-\beta|01\rangle+\gamma|10\rangle-\delta|11\rangle$ & $I\otimes Z$ 
            & \multirow{4}{*}{1100} & 00 & $\alpha|00\rangle-\beta|01\rangle-\gamma|10\rangle+\delta|11\rangle$ & $Z\otimes Z$ \\
            & 01 & $\alpha|00\rangle+\beta|01\rangle+\gamma|10\rangle+\delta|11\rangle$ & $I\otimes I$ & 
            & 01 & $\alpha|00\rangle+\beta|01\rangle-\gamma|10\rangle-\delta|11\rangle$ & $Z\otimes I$ \\
            & 10 & $\alpha|00\rangle-\beta|01\rangle-\gamma|10\rangle+\delta|11\rangle$ & $Z\otimes Z$ & 
            & 10 & $\alpha|00\rangle-\beta|01\rangle+\gamma|10\rangle-\delta|11\rangle$ & $I\otimes Z$ \\
            & 11 & $\alpha|00\rangle+\beta|01\rangle-\gamma|10\rangle-\delta|11\rangle$ & $Z\otimes I$ & 
            & 11 & $\alpha|00\rangle+\beta|01\rangle+\gamma|10\rangle+\delta|11\rangle$ & $I\otimes I$ \\
            \midrule
            \multirow{4}{*}{0101} & 00 & $-\alpha|01\rangle+\beta|00\rangle-\gamma|11\rangle+\delta|10\rangle$ & $I\otimes ZX$ 
            & \multirow{4}{*}{1101} & 00 & $-\alpha|01\rangle+\beta|00\rangle+\gamma|11\rangle-\delta|10\rangle$ & $Z\otimes ZX$ \\
            & 01 & $-\alpha|01\rangle-\beta|00\rangle-\gamma|11\rangle-\delta|10\rangle$ & $I\otimes ZXZ$ & 
            & 01 & $-\alpha|01\rangle-\beta|00\rangle+\gamma|11\rangle+\delta|10\rangle$ & $Z\otimes ZXZ$ \\
            & 10 & $-\alpha|01\rangle+\beta|00\rangle+\gamma|11\rangle-\delta|10\rangle$ & $Z\otimes ZX$ & 
            & 10 & $-\alpha|01\rangle+\beta|00\rangle-\gamma|11\rangle+\delta|10\rangle$ & $I\otimes ZX$ \\
            & 11 & $-\alpha|01\rangle-\beta|00\rangle+\gamma|11\rangle+\delta|10\rangle$ & $Z\otimes ZXZ$ & 
            & 11 & $-\alpha|01\rangle-\beta|00\rangle-\gamma|11\rangle-\delta|10\rangle$ & $I\otimes ZXZ$ \\
            \midrule
            \multirow{4}{*}{0110} & 00 & $\alpha|10\rangle-\beta|11\rangle+\gamma|00\rangle-\delta|01\rangle$ & $X\otimes Z$ 
            & \multirow{4}{*}{1110} & 00 & $-\alpha|10\rangle+\beta|11\rangle+\gamma|00\rangle-\delta|01\rangle$ & $ZX\otimes Z$ \\
            & 01 & $+\alpha|10\rangle+\beta|11\rangle+\gamma|00\rangle+\delta|01\rangle$ & $X\otimes I$ & 
            & 01 & $-\alpha|10\rangle-\beta|11\rangle+\gamma|00\rangle+\delta|01\rangle$ & $ZX\otimes I$ \\
            & 10 & $\alpha|10\rangle-\beta|11\rangle-\gamma|00\rangle+\delta|01\rangle$ & $XZ\otimes Z$ & 
            & 10 & $-\alpha|10\rangle+\beta|11\rangle-\gamma|00\rangle+\delta|01\rangle$ & $ZXZ\otimes Z$ \\
            & 11 & $\alpha|10\rangle+\beta|11\rangle-\gamma|00\rangle-\delta|01\rangle$ & $XZ\otimes I$ & 
            & 11 & $-\alpha|10\rangle-\beta|11\rangle-\gamma|00\rangle-\delta|01\rangle$ & $ZXZ\otimes I$ \\
            \midrule
            \multirow{4}{*}{0111} & 00 & $-\alpha|11\rangle+\beta|10\rangle-\gamma|01\rangle+\delta|00\rangle$ & $X\otimes ZX$ 
            & \multirow{4}{*}{1111} & 00 & $\alpha|11\rangle-\beta|10\rangle-\gamma|01\rangle+\delta|00\rangle$ & $ZX\otimes ZX$ \\
            & 01 & $-\alpha|11\rangle-\beta|10\rangle-\gamma|01\rangle-\delta|00\rangle$ & $X\otimes ZXZ$ & 
            & 01 & $\alpha|11\rangle+\beta|10\rangle-\gamma|01\rangle-\delta|00\rangle$ & $ZX\otimes ZXZ$ \\
            & 10 & $-\alpha|11\rangle+\beta|10\rangle+\gamma|01\rangle-\delta|00\rangle$ & $XZ\otimes ZX$ & 
            & 10 & $\alpha|11\rangle-\beta|10\rangle+\gamma|01\rangle-\delta|00\rangle$ & $ZXZ\otimes ZX$ \\
            & 11 & $-\alpha|11\rangle-\beta|10\rangle+\gamma|01\rangle+\delta|00\rangle$ & $XZ\otimes ZXZ$ & 
            & 11 & $\alpha|11\rangle+\beta|10\rangle+\gamma|01\rangle+\delta|00\rangle$ & $ZXZ\otimes ZXZ$ \\
            \bottomrule
        \end{tabular}
    \end{table}
\end{landscape}
\subsection{Control of teleportation of \textit{n}-qubits via \textit{m} participants}
\label{sec_3.2}
Increasing the number of participants in the protocol by one reveals a recurring pattern: an additional Z gate is applied to the end of Bob's qubits, consistent with Equation \ref{eqn_20}.
The message state is an $n$-qubit state $|\psi^n\rangle$ given below in Equation \ref{eqn_28}.
\begin{equation}
    \label{eqn_28}
    |\psi^n\rangle= \sum_{j \in\{0,1\}^n} \alpha_j |j\rangle
\end{equation}
where $\sum^{2^n-1}_{j=0} |\alpha_j|^2 = 1$. The state $|\psi^n\rangle$ can be entangled depending on the coefficients $\alpha_j$. The teleportation resource is an $(m \times n)$-qubit state, formed by $n$ instances of $m$-qubit GHZ states, each defined by Equation \ref{eqn_16}. The total resource state is given in Equation \ref{eqn_30}. The $m$ participants have $n$ resource qubits each.
\begin{equation}
    \label{eqn_30}
    |n,m\text{GHZ}\rangle=\frac{1}{\sqrt{2^n}} \left[
    \begin{aligned}
        (|0\rangle^{\otimes m} &+ |1\rangle^{\otimes m})_{q_nq_{2n}q_{3n}..q_{mn}} \\
        (|0\rangle^{\otimes m} &+ |1\rangle^{\otimes m})_{q_{n+1}q_{2n+1}q_{3n+1}..q_{mn+1}} \\
        &\vdots \\
        (|0\rangle^{\otimes m} &+ |1\rangle^{\otimes m})_{q_{2n-1}q_{3n-1}q_{4n-1}..q_{(m+1)n-1}} \\
    \end{aligned}
    \right]
\end{equation}
The composite system state is given in Equation \ref{eqn_31}.
\begin{align}
    \label{eqn_31}
    |\psi^n\rangle_{q_0..q_{n-1}} \otimes &|n,m\text{GHZ}\rangle_{q_n..q_{(m+1)n-1}} \nonumber \\
    &=\frac{1}{\sqrt{2^n}} \left[ \sum_{i \in \{0,1\}^n} \alpha_i |i\rangle \otimes \left\{
    \begin{aligned}
        (|0\rangle^{\otimes m} &+ |1\rangle^{\otimes m})_{q_nq_{2n}q_{3n}..q_{mn}} \\
        (|0\rangle^{\otimes m} &+ |1\rangle^{\otimes m})_{q_{n+1}q_{2n+1}q_{3n+1}..q_{mn+1}} \\
        &\vdots \\
        (|0\rangle^{\otimes m} &+ |1\rangle^{\otimes m})_{q_{2n-1}q_{3n-1}q_{4n-1}..q_{(m+1)n-1}}
    \end{aligned}
    \right\} \right]
\end{align}
Alice performs BMs on pairs of her qubits, each pair consisting of a message qubit and a corresponding qubit from the resource state, and saves her measurement results in classical bits $c_0,..,c_{2n-1}$. The BM operators that Alice has to apply to her corresponding qubits are given in Equation \ref{eqn_32}.
\begin{equation}
    \label{eqn_32}
    \prod^{n-1}_{i=0}BM_{q_i, q_{(i+n)}}=BM_{q_0,q_n}\cdot BM_{q_1,q_{n+1}}\cdot ... \cdot BM_{q_{n-1},q_{2n-1}}
\end{equation}

The intermediate participants $\text{Charlie}_i$ each apply a Hadamard gate on their qubits and measure them in the Z basis, saving the results in classical bits $c_{2n},..,c_{mn-1}$. The parties then send all classical bits to Bob, using which he applies UR to his qubits. Bob possesses $n$ qubits, viz., $q_{mn},q_{mn+1},..,q_{mn+n-1}$. These qubits can be labeled using an integer index $j$, as given below in Equation \ref{eqn_33}.
\begin{equation}
    \label{eqn_33}
    \text{Bob's j$^{th}$ qubit} \implies q_{mn+j}
\end{equation}
where $j$ varies from 0 to $(n-1)$. Bob's sequence of UR to each of his qubits are summarized below in Equation \ref{eqn_34}.
\begin{equation}
    \label{eqn_34}
    \text{Operators applied on j$^{th}$ qubit }(q_{mn+j}) \implies Z^{c_j}\cdot X^{c_{(n+j)}}\cdot \prod^{m-1}_{i=2}Z^{c_{(in+j)}}
\end{equation}
\begin{figure}
    \centering
    \includegraphics[width=\columnwidth]{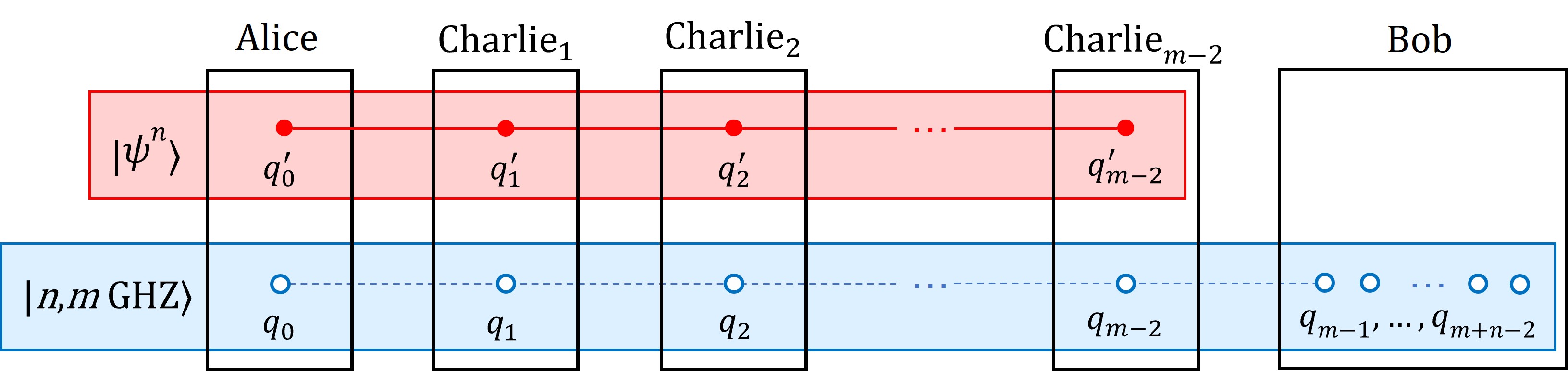}
    \caption{\label{fig_5}Schematic representation of the protocol in the distributed message state and minimal resource configurations, illustrating the allocation of qubits among participants.}
\end{figure}

\section{Results and Discussion}
\label{sec_5}
IBM's Qiskit SDK has been used to implement our protocol for various $m$ and $n$ values using the quantum computer simulator \textit{AerSimulator}. All messages states for the protocol were arbitrarily generated.

\begin{figure}
    \centering
    \includegraphics[width=\columnwidth]{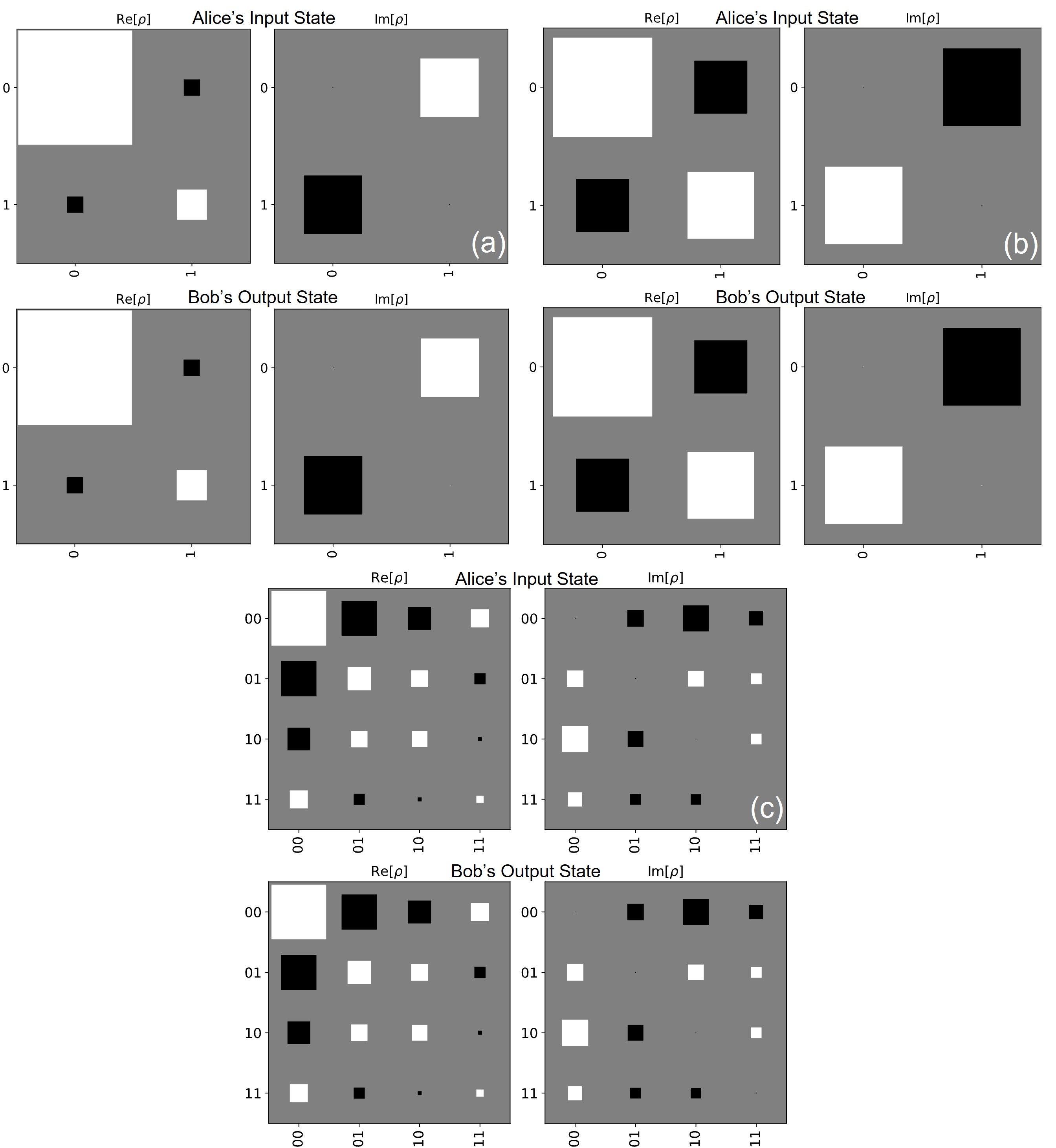}
    \caption{\label{fig_4}(a) Hinton maps for Alice's input state $|\psi^{1}_{in}\rangle$ and Bob's output state $|\psi^{1}_{out}\rangle$ for 3-participant teleportation. (b) Hinton maps for input and output states for 4-participant teleportation. (c) Hinton maps for Alice's input state $|\psi^{2}_{in}\rangle$ and Bob's output state $|\psi^{2}_{out}\rangle$ for 3-participant 2-qubit teleportation. The close resemblance between the Hinton diagrams for corresponding input and output states verify the observed teleportation fidelity of $\approx1$.}
\end{figure}

For 1-qubit Controlled Teleportation, the input state $|\psi^{1}_{in}\rangle$ can be written as its density matrix operator $\rho^{1}_{in}$ by taking the self-outer product of the state, given below in Equation \ref{eqn_36}.
\begin{equation}
    \label{eqn_36}
    \rho^{1}_{in} = |\psi^{1}_{in}\rangle\langle\psi^{1}_{in}|
\end{equation}
$\rho^{1}_{in}$ is a $2\times2$ matrix having 4 complex elements. We choose to represent density matrices as Hinton diagrams, given in Figure \ref{fig_4}. A Hinton diagram is a visual tool that depicts the magnitude and sign of a density matrix. The size of each square in the diagram is proportional to the magnitude of the corresponding matrix element, and the colour of the squares is used to represent their sign. In Figure \ref{fig_4}, the left diagram represents real part of the matrix elements, while the right diagram represents their imaginary parts. For higher-dimensional quantum systems involving multiple qubits, a Hinton diagram helps to quickly identify the significant elements large matrices. In Figure \ref{fig_4} (a) \& (b), the Hinton maps represent 1-qubit states, i.e., $2\times2$ density matrices, while in Figure \ref{fig_4} (c), the Hinton map represents a 2-qubit state, i.e., a $4\times4$ density matrix.

Following Bob's UR in the protocol, he obtains the 1-qubit output state $|\psi^{1}_{out}\rangle$. The density matrix $\rho^{1}_{out}$ is constructed in similar fashion by taking the self-outer product of $|\psi^{1}_{out}\rangle$, given below in Equation \ref{eqn_37}.
\begin{equation}
    \label{eqn_37}
    \rho^{1}_{out} = |\psi^{1}_{out}\rangle\langle\psi^{1}_{out}|
\end{equation}
We calculate the fidelity between the input and output density matrices to verify faithful teleportation using the formula given below in Equation \ref{eqn_38}.
\begin{equation}
    \label{eqn_38}
    F(\rho_{in}, \rho_{out}) = Tr\bigg(\sqrt{\sqrt{\rho_{in}}\cdot\rho_{out}\cdot\sqrt{\rho_{in}}}\bigg)^2
\end{equation}

If any of the intermediate participants turn unfaithful and does not to forward their measurement results to the end receiver, the reconstruction of the unknown quantum state $|\psi^n\rangle$ will no longer be deterministic. Hence each intermediate participant $\text{Charlie}_i$ acts as a controller, making this protocol operate as a \textit{multi-controller} teleportation scheme. Further, the protocol can be used in various configurations depending on the distribution of the resource qubits and the message qubits, maximizing overall security and quantum resource utilization of the protocol:

\textit{Message State Distribution}: The $n$ qubits of the message state $|\psi^n\rangle$ can be distributed unevenly among Alice and intermediate participants, i.e., Alice does not need to have all $n$ message qubits. This configuration change enhances security as it mitigates the risk of interception by an eavesdropper Eve, seeking to steal Alice's message qubits using a targeted attack. Due to this configuration, Eve could only capture a fraction of the message qubits from a single participant, rendering the information gained useless without access to all message qubits. The protocol is modified such that the participants that receive the message qubits perform BMs on pairs of qubits consisting of one message qubit and one resource qubit. For resource qubits not included in these pairs and for participants that do not receive any message qubits, they apply H gates and measure in the Z basis. All measurements results are then sent to the receiver Bob, using which he applies UR on each of his qubits.

\textit{Minimal Resource Usage}: The qubits of the teleportation resource state can be distributed such that Alice and all intermediate participants have one resource qubit each, while the end receiver has $n$ qubits. This configuration minimizes the teleportation resource usage of the protocol without affecting its operation, since number of resource qubits required is $(m+n-1)$ instead of the typical $(m\times n)$.

\textit{Selective End Receiver}: Alice and the intermediate parties can collaborate and dynamically select the participant who becomes the end receiver Bob during the protocol execution by choosing to whom Alice sends her $q_1$ measurement results, i.e., $c_1$. Once Alice selects the end receiver, she must notify the intermediate participants of her selection. Until this disclosure, the end receiver's identity remains unknown to all participants, enhancing the protocol's security by keeping the end receiver incognito. This prevents unfaithful participants from revealing the end receiver and making them susceptible to a potential eavesdropper Eve's interception attacks.

\section{Conclusion}
\label{sec_6}

The proposed protocol establishes a framework for teleporting multi-qubit states across a distributed network of $m$ participants, using $n$ instances of $m$-qubit GHZ states as the quantum teleportation resource. On top of this, miscellaneous configurations of the protocol are introduced which increase the overall security and quantum resource utilization of the protocol. Future work on configurations for other protocols such as quantum dialogue, which can increase the practical implementation costs of distant quantum computing networks.

Future work on analysis on noise resilience will be essential to assess and enhance the protocol's robustness. Evaluating fidelity in the context of real quantum computing hardware will help identify and address areas of increased susceptibility to noise. Understanding how noise impacts specific operations and identifying segments of the protocol that are particularly error-prone will enable improvements in overall teleportation fidelity during the implementation of the protocol.

\section*{Acknowledgments}
We would like to express our sincere gratitude to Mr. Abhijit Hazra for his fruitful discussions and suggestions during the course of our research work. We acknowledge the support in part by the Ministry of Electronics and Information Technology (MeitY), Quantum Computing Applications Laboratory (QCAL), Government of India, which granted access to Amazon Braket, a cloud quantum computing facility through Amazon Web Services (AWS). We acknowledge IBM Quantum Lab software interface and IBM Qiskit library and, which provided the platform to simulate our teleportation protocol circuit and gave us insights on our work.

\section*{Ethics Statements}
\textit{Conflict of Interest}: There is no conflict of interests to declare among the authors of this manuscript.

\noindent \textit{Funding}: The authors declare that no funds, grants or other financial support was received during the preparation of this manuscript.

\noindent \textit{Data Availability}: No datasets were generated or analysed during the current study.

\end{document}